\documentclass[5p,authoryear]{elsarticle}

\usepackage{graphicx}

\usepackage{amsmath,amssymb}
\usepackage{hyperref}
\usepackage{appendix}
\usepackage{bm}
\hypersetup{ 
colorlinks,%
citecolor=blue,
filecolor=blue,
linkcolor=blue,
urlcolor=blue 
}

\journal{Journal of Theoretical Biology}

\begin{document}

\begin{frontmatter}

\title{Predation effects on mean time to extinction under demographic stochasticity}

\author[gm]{Gian Marco Palamara}

\author[gd]{Gustav W. Delius}
\address[gd]{Department of Mathematics, The University of York, York, United Kingdom}

\author[ms]{Matthew J. Smith}
\address[ms]{Computational Ecology and Environmental Science Group, Microsoft Research, Cambridge, United Kingdom}

\author[op]{Owen L. Petchey}
\address[gm, op]{Institute of Evolutionary Biology and Environmental Studies, The Univiersity of Zurich, Zurich, Switzerland}

\begin{abstract}
Methods for predicting the probability and timing of a species' extinction are typically based on a combination of theoretical models and empirical data, and focus on single species population dynamics. Of course, species also interact with each other, forming more or less complex networks of interactions. Models to assess extinction risk often lack explicit incorporation of these interspecific interactions.
We study a birth and death process in which the death rate includes an effect from predation. This predation rate is included via a general nonlinear expression for the functional response of predation to prey density. We investigate the effects of the foraging parameters (e.g. attack rate and handling time) on the mean time to extinction.

Mean time to extinction varies by orders of magnitude when we alter the foraging parameters, even when we exclude the effects of these parameters on the equilibrium population size. In particular we observe an exponential dependence of the mean time to extinction on handling time. These findings clearly show that accounting for the nature of interspecific interactions is likely to be critically important when estimating extinction risk.
\end{abstract}

\begin{keyword}

Trophic Interaction \sep Predator Prey Model \sep Birth and Death Process \sep Quasistationary Distribution 

\end{keyword}

\end{frontmatter}

\section{Introduction}

Population ecologists have long sought to understand the relationship between the probability that a population will go extinct, exogenous factors (e.g. environmental variability) and endogenous factors (e.g. body size, life history, trophic position) \citep{Lande2004}. As well as providing information useful for planning conservation of individual species, understanding these relationships promises insights into the determinants of ecosystem stability and complexity \citep{McCann2000, Allesina2012}.

Populations can take different routes to extinction. Extinction can be observed as a gradually declining population size due to habitat deterioration or as  a crash due to chance events like random catastrophes or random fluctuations in population size (demographic stochasticity) \citep{Lande1993}. Demographic stochasticity is caused by random variation in individual mortality and reproduction and is independent from time. Environmental stochasticity is, on the other hand, random variation in time of individual rates \citep{Lande2004}. While environmental stochasticity can be important for populations of any size, demographic stochasticity becomes particularly important at low population sizes.

Classical population theory shows that demographic stochasticity causes the mean time to extinction to increase exponentially with equilibrium population size, while variation in environmental conditions can lead to a power law scaling \citep{Lande1993, Foley1994}. Ecologists have shown which biological features affect species' risk of extinction. Factors such as slow life history and small geographical range size are all independently associated with a high extinction risk \citep{Purvis2000}. Other effects such as genetic and environmental change have been included into this species-centric approach \citep{Purvis2008, Lee2011}. One outcome of this large body of research is population viability analysis (PVA) \citep{Brook2000, Beissinger2002}. A related outcome is the classification of species into extinction risk categories \citep{Mace2008}. However single species models used for assessing population's viability often lack explicit incorporation of direct trophic interactions\citep{Sabo2007, Sabo2008}.

Interspecific interactions have been widely studied theoretically and experimentally in the fields of population and community ecology. The pioneering work of Holling \citep{Holling1959} led to a non-linear density dependent interaction  term called the \emph{Functional Response of Predation} and the effect of functional response on the stability of communities has been widely studied \citep{Drossel2004}. There have been several modifications to the original Holling's functional response, describing in detail different foraging mechanisms \citep{Real1977, Abrams2000, Jeschke2002}. It has been shown how the introduction of a scaling exponent between predator attack rate and prey density can drastically alter the dynamics of realistic systems. This scaling exponent stabilizes chaotic dynamics and reduces or eliminates extinctions in deterministic food web models \citep{Williams2004}.

Interspecific interactions could also influence extinction time. For example, predation pressure could reduce population size and thereby increase the chance of extinction via variation in population size resulting from demographic stochasticity. Predators with different foraging behaviors may reduce or increase the risk of extinction of their prey according to their feeding strategies. Most PVAs assume that any important effects of interspecific interactions are summarized in population level parameters \citep{Sabo2007}, for example, predation is included as a constant (density independent) source of mortality rather than a coupled, density dependent population process \citep{Sabo2008}. Single species models used in PVA often fail to separate stochastic variation from population cycles induced by species interaction, and this can bias forecasts of viability \citep{Sabo2007, Sabo2008}.

In this paper we present a single species model in which the demographic effects of predation on small prey populations are explicitly investigated. We aim to clarify the effects of interspecific interactions on the extinction process. We analyze a single species birth-death process in which the death rate includes density dependent mortality in the form of predation by a fixed abundance of predators. We then investigate the importance of the predation functional response parameters on the mean time to extinction. This investigation is novel in itself, and also represents an initial step towards a more complete appreciation of the effects of interspecific interactions on time to extinction. We use the single species framework in order to gain insight into the behavior of more complex models.

\section{Stochastic and deterministic models} \label{mm}

We describe the dynamics of single species population with number of individuals $n(t)$ at time $t$ by defining functions for birth rate $b(n)$ and death rate $d(n)$ as
\begin{eqnarray}
b(n) &=& \lambda \left(1-\frac{n}{k}\right), \nonumber \\
d(n) &=& \mu + f(n)p, \label{bd} 
\end{eqnarray}
The logistic form of the birth rate function represents density dependent effects such as intraspecific competition for resources. The parameter $\lambda$ is the per capita birth rate of the prey species and $k$ represents the maximum number of individuals the environment can support \citep{Nisbet1982, Nasell2001}. The death rate includes a constant term $\mu$, the predator-free per capita death rate, and the term $f(n)p$, the per capita functional response of predation rate to prey abundance. The function $p(t)$ is the number of predator individuals at time $t$. The per capita growth rate of the population is then the difference between per capita birth and death rates i.e., $r=\lambda - \mu$.

The functional response $f(n)$ specifies the effect of prey density on the predation rate. Historically several functional responses have been proposed to describe this trophic interaction \citep{Jeschke2002}. We used the general expression of the type III functional response
\begin{equation}
f(n) = \frac{\alpha n^q}{1 + h\alpha n^{q+1}}, \label{fr}
\end{equation}
where $\alpha$ is the attack rate (a measure of the encounter rate and capture success of the predator foraging on the prey), $h$ is the handling time (a measure of the time needed to attack, eat and digest the prey) and $q$ is a scaling exponent between encounter rate and prey density. Different values of $q$ are generally assumed to represent different types of predation. The type III functional response has been associated with learning effects of the predator in catching and handling its prey \citep{Real1977}, with generalist predators switching among alternate prey \citep{Smout2010} or with spatial effects enabling prey to hide from predators \citep{Vucic-Pestic2010}. Setting $q=0$ into expression \eqref{fr} we obtain the type II functional response and setting $q=0$ and $h=0$ we obtain the type I functional response \citep{Holling1959, Real1977}.

The range of variation of the foraging parameters is constrained by biological arguments:
\begin{itemize}
\item Handling time $h$ takes values between $0.001 days$ and $0.1 days$. We assume that $h$ cannot be larger than the average lifetime of the prey species (assumed to be $1/\lambda$).

\item Attack rate $\alpha$ takes values between $0.001 day^{-1} ind^{-1}$ and $10 day^{-1} ind^{-1}$. This choice is justified by empirical observations of foraging behavior of insects and other organisms \citep{Vucic-Pestic2010, Hammill2010}.

\item The exponent $q$ takes values between 0 and 2 again due to empirical observations \citep{Vucic-Pestic2010} and theoretical work \citep{Williams2004}.
\end{itemize}

Combining expressions \eqref{bd} and \eqref{fr} we define the population birth and death rates $B(n)$ and $D(n)$ as 
\begin{eqnarray}
B(n) &=& nb(n) = n\lambda \left(1-\frac{n}{k}\right), \nonumber \\
D(n) &=& nd(n) = n\mu + \frac{\alpha pn^{q+1}}{1+\alpha hn^{q+1}}. \label{BD} 
\end{eqnarray}
The state of the system is completely characterized by the probability $p(n,t)$ of having $n$ individuals at time $t$, where $n$ takes integer values in the range $\{0, \cdots, k\}$. The master equation describing the time evolution of this resulting probability distribution is
\begin{equation}
\begin{split}
\frac{dp(n,t)}{dt}=&D(n+1)p(n+1,t) + B(n-1)p(n-1,t) \\
&- (B(n)+D(n))p(n,t). \label{me}
\end{split}
\end{equation}
In order to express equation \eqref{me} in a compact way we introduce the conditions $p(k+1,t)=p(-1,t)=0$. Once we know the initial condition $p_0(n)=p(n,0)$, then the state probabilities are given by the solution of equation \eqref{me}.
The process represented in this master equation has a degenerate stationary distribution $\overline{{\bf p}}=(1,0,\cdots,0)$, which the distribution $p(n,t)$ approaches as time $t$ approaches infinity. In other words, extinction is ultimately a certainty. However, in section \ref{stcb} we will show how to compute the conditioned probability of having a population of size $n$ at time $t$ conditioned on the fact that it has not yet gone extinct. In order to understand the qualitative behavior of the model it also helps to understand the dynamics of the associated rate equation.

\subsection{Deterministic behavior}

Every stochastic birth and death process has an associated  deterministic rate equation, describing the time evolution of the mean population size \citep{Gardinier2009}, given by
\begin{equation}
\frac{dn}{dt}= B(n)-D(n). \label{detrat}
\end{equation}
In our case, substituting the rates \eqref{BD} into equation \eqref{detrat} we obtain
\begin{equation}
\frac{dn}{dt}= \lambda n\left(1-\frac{n}{k}\right) - \mu n - \frac{\alpha pn^{q+1}}{1+\alpha h n^{q+1}}. \label{det}
\end{equation}
At this point we assume constant presence of predators and set the number of predators $p$ as a constant. This choice enable us to investigate the effect of a generalist predator whose density is not affected by the prey density. In order to understand how the prey population goes extinct, in appendix \ref{App1} we derive an adimensional formulation of equation \eqref{det} and compute all its fixed points and their stability.

\begin{figure*}
  \begin{center}
    \includegraphics[width=0.9\textwidth]{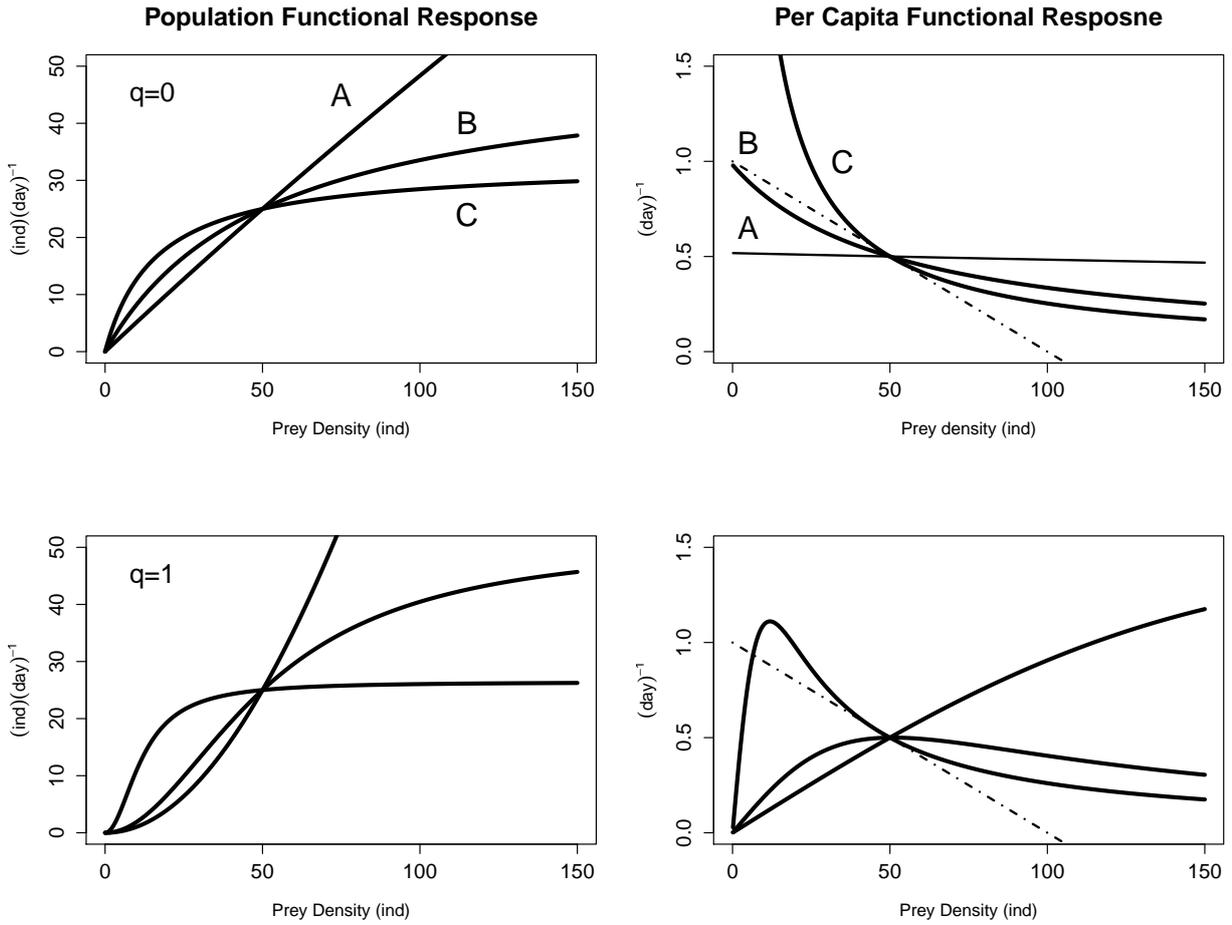}
  \end{center}
\caption{Prey population mortality due to predation (left panels) and per capita mortality due to predation (right panels) following a type II (upper panels) and a type III (lower panels) functional response. On the left plots the continuous lines A, B and C represent different population functional responses. On the right plots solid lines are the per capita functional response while the dashed line is the per capita growth curve ($\lambda = 1.5 \, day^{-1}$; $\mu = 0.5 \, day^{-1}$ and $k = 150 \, ind$). The number of predators is fixed at $p=1 \, ind$. The foraging parameters are A: $h = 0.001 \,days$; $\alpha = 0.5 \, day^{-1}ind^{-1}$. B: $h = 0.02 \, days$; $\alpha = 1 \, day^{-1}ind^{-1}$ at the transcritical bifurcation. C: $h = 0.03 \, days$; $\alpha = 2.05 \, day^{-1}ind^{-1}$. Note that with this particular choice of foraging parameters the value of the fixed point is fixed at $n0 = 50 \, ind$.}\label{fig1}
\end{figure*}

In the deterministic rate equation \eqref{det} the extinction state $n=0$ is always a fixed point, while the other fixed points are given by the intercepts between the per capita growth term $(r-\lambda n/k)$ and the per capita functional response $pf(n)$ (see figure \ref{fig1}). Two different extinction scenarios are revealed by the analysis of the deterministic equation, associated with changes in the stability of the extinction state \citep{Assaf2010}. When $q=0$ (type II functional response) the model has at most two non zero fixed points and the extinction state can be stable or unstable. When $q > 0$ (type III functional response) the extinction state will always be unstable.

For the type II functional response the extinction state is stable if $p\alpha > r$. In this case there are two positive fixed points, one stable and one unstable (see figure \ref{fig1}). At $p\alpha = r$ there is a transcritical bifurcation, at which the unstable fixed point goes to 0 (see figure \ref{fig1}) and the extinction state becomes unstable. When the extinction state is unstable ($p\alpha < r$) there is only one other fixed point and this is stable (see figure \ref{fig1}). For the type III functional response ($q>0$) the extinction state will always be unstable like in the case of the type II functional response after the transcritical bifurcation. But, differently from the type II functional response, for $q>0$ there can be up to three non zero fixed points (see appendix \ref{App1}). In this case there are combinations of foraging parameters which lead to two stable fixed points (see figure \ref{fig1}).

\subsection{Stochastic behavior} \label{stcb}

Given the birth and death process (equation \eqref{BD}), extinction is caused by demographic stochasticity and typically occurs in two different ways depending on whether the extinction state is a stable or unstable fixed point of the deterministic rate equation \eqref{det}. For the type II functional response, before the bifurcation, extinction is caused by a large fluctuation which brings the system from the stable to below the unstable fixed point. From there the fast deterministic evolution takes the population quickly to the stable extinction state. For the type II functional response after the bifurcation, and for the type III functional response, extinction is caused by a large fluctuation in density which brings the system from the stable fixed point to the unstable, absorbing extinction state. For the type III functional response, when there are two stable fixed points, extinction needs two large fluctuations to occur, one fluctuation which brings the system from the larger to the lower stable fixed point and another fluctuation which brings the system from the lower stable fixed point to the unstable extinction state. This result confirms the stabilizing effect of the type III functional response \citep{Williams2004}.

When the deterministic rate equation has at least one stable fixed point, the system approaches a quasi stationary state with a time independent distribution $\pi(n)$; this is called the \emph{Quasistationary Distribution} \citep{Bartlett1960, Nisbet1982}. The quasistationary distribution $\pi(n)$ is obtained from the probability $p_c(n,t)$ of finding $n$ individuals at time $t$, conditioned on the fact extinction has not occurred yet:
\begin{equation}
p_c(n,t)=\frac{p(n,t)}{1-p(0,t)}. \label{pc}
\end{equation}
We derive a master equation for the conditioned probability $p_c(n,t)$ and look for its stationary solution $\pi(n)$ (see appendix \ref{App2}). When the initial condition of equation \eqref{me} is set to the quasistationary distribution, then the probability of finding $n$ individuals at time $t$ becomes
\begin{equation}
p(n,t) \simeq \pi(n)\exp(-t/MTE). \label{pi}
\end{equation}
The time to extinction is then an exponentially distributed random variable with mean equal to $MTE$ and, if the initial condition of \eqref{me} is set equal to $\pi(n)$, then the mean time to extinction is
\begin{equation}
MTE = \frac{1}{D(1)\pi(1)}. \label{mte}
\end{equation}
There are more complicated expressions for the mean time to extinction when the initial condition is not the quasi stationary distribution $\pi(n)$ (see appendix \ref{App2}).

We compute the $MTE$ of the birth and death process \eqref{BD} for different functional responses and for different values of the foraging parameters within the functional responses. All other parameters remain fixed.

\subsection{Numerical calculations} 

Despite its apparent simplicity it is not possible to obtain closed expressions for the quasistationary distribution $\pi(n)$ of the birth and death process \eqref{BD}. Instead we obtain the quasistationary distribution in a realistic range of foraging parameters by an iterative numerical scheme described in appendix \ref{App2}.
In order to perform numerical calculations we fix the growth parameters of equation \eqref{det} in the following way: $\lambda = 1.5 \, day^{-1}$; $\mu = 0.5 \, day^{-1}$; $k=150 \, ind$. With that choice the intrinsic growth rate of the prey population is fixed to $1 \, day^{-1}$. If we look at equation \eqref{det} without predators i.e., putting $p=0$, we obtain a modified version of the classic logistic equation:
\begin{equation}
\frac{dn}{dt}=rn - \frac{\lambda n^2}{k}. \label{unpert}
\end{equation}
The modified logistic equation \eqref{unpert} has a non zero fixed point $n^*=kr/\lambda = 100 \, ind$. The system behaves according to equation \eqref{unpert} when predation is absent or very weak.

We fix the number of predators to $p= 1\, ind$ and compute the logarithm of the Mean Time to Extinction (MTE) of the birth and death process \eqref{BD} obtained using the quasi stationary distribution as initial condition (see appendix \ref{App2}) for different combinations of foraging parameters. In the case of the type II functional response ($q=0$), we also compute the MTE using different choices of initial conditions. In this case the time needed to relax to the quasi stationary distribution is negligible with respect to the MTE.

In order to quantify the difference between the effects of the foraging parameters on the equilibrium population density and the direct effect of the foraging parameters  on the mean time to extinction we also perform numerical simulations of the model changing the foraging parameters and keeping fixed the stable fixed point. As we increase the handling time, we also increase the attack rate in order to keep the effective predation pressure constant at the equilibrium population. Using the equilibrium relation $B(n_0)=D(n_0)$, we find a relation between attack rate and handling time which includes one fixed point $n_0$ as another parameter
\begin{equation}
\alpha_q = \frac{\alpha(h,n_0)}{n^q_0} = \frac{rk - \lambda n_0}{n_0^{q}[pk - hn_0(rk - \lambda n_0)]}. \label{ah}
\end{equation}
In order to avoid fixing an unstable equilibrium with relation \eqref{ah}, we limit our investigation to those values of handling time which give raise to a negative Jacobian (see appendix \ref{App1}) i.e.,
\begin{equation}
h < h_1 = \frac{\lambda pk [ (1-q) \lambda n_0 + qrk]}{n_0(1+q)(rk - \lambda n_0)^2}. \label{hh}
\end{equation}
As the handling time approaches the value
\begin{equation}
h_0 = \frac{pk}{n_0(rk-\lambda n_0)}, \label{h0}
\end{equation} 
the required attack rate approaches infinity, so we limit ourselves to $h<min(h_0,h_1)$. In the case of $q>0$ we also keep $h$ small enough so that we do not enter the bistable region. Matlab code to reproduce our calculations can be found at \url{http://purl.org/net/extinction_code}.

\section{Results}

\begin{figure}
  \begin{center}
    \includegraphics[width=0.5\textwidth]{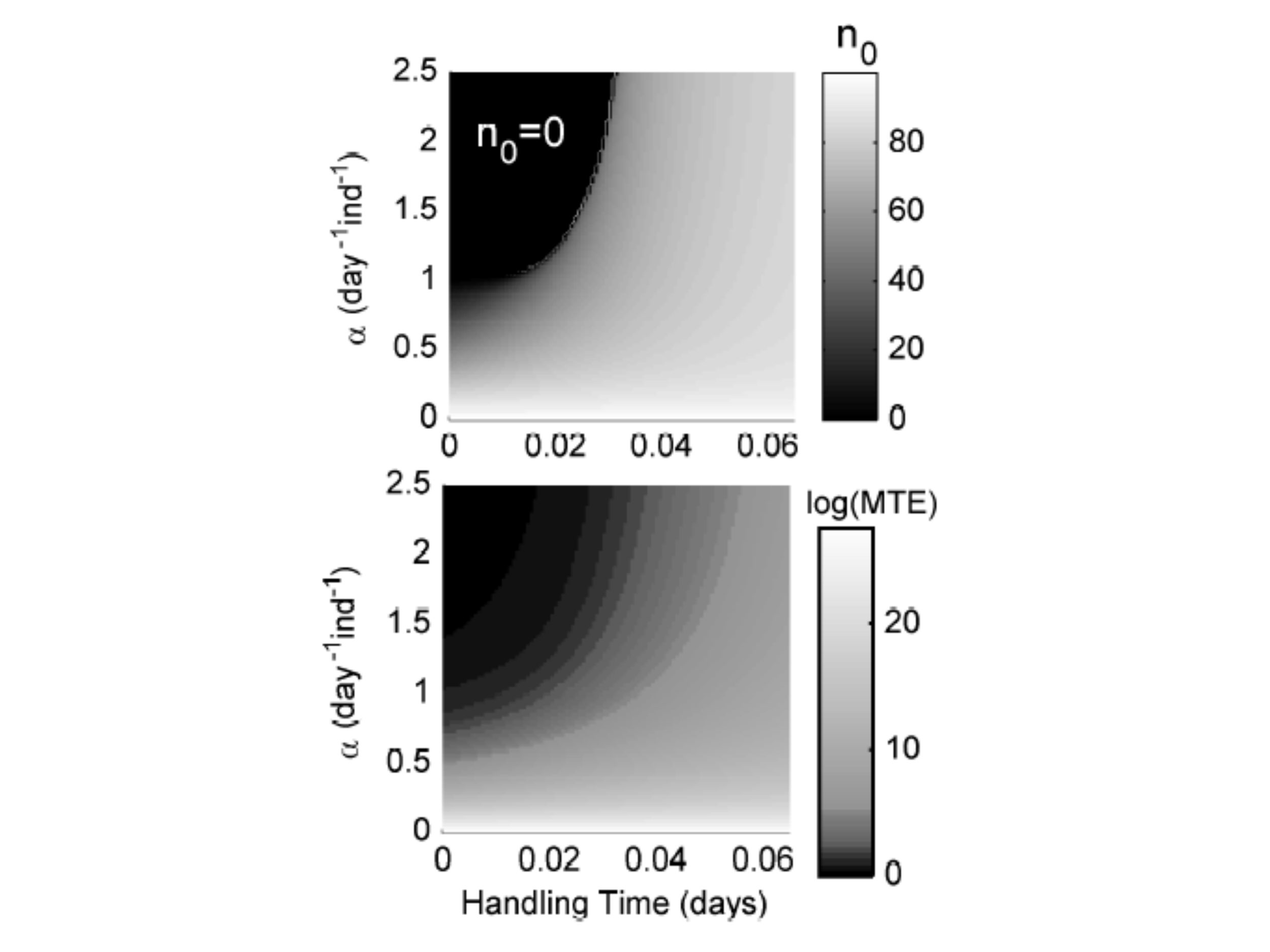}
  \end{center}
\caption{Variation of the stable fixed point $n_0$ predicted by equation \eqref{det} (upper panel) and the log of the mean time to extinction (MTE) of the birth and death process \eqref{bd} using the quasistationary distribution as initial condition (lower panel) as a function of handling time and attack rate for type II functional response. The growth parameters and the number of predators are as specified in the legend of figure \ref{fig1}.}\label{fig2}
\end{figure}

The foraging parameters of the functional response affect the MTE in direct and indirect ways. The foraging parameters affect the MTE indirectly by influencing the equilibrium population size of the prey. They influence it directly by changing the MTE even for a fixed equilibrium prey population size. The indirect effects of the foraging parameters on MTE can be seen in figure \ref{fig2}. As expected, the MTE is relatively low when the functional response parameters lead to extinction being predicted by the deterministic model (region labeled $n_0 = 0$ in figure \ref{fig2}). When the foraging parameters are such that there is a stable positive equilibrium, the MTE is clearly positively, but non linearly, related to the equilibrium population size. Changes in both attack rate and handling time strongly influence MTE through changes in equilibrium population size.

\begin{figure*}
  \begin{center}
    \includegraphics[width=0.65\textwidth]{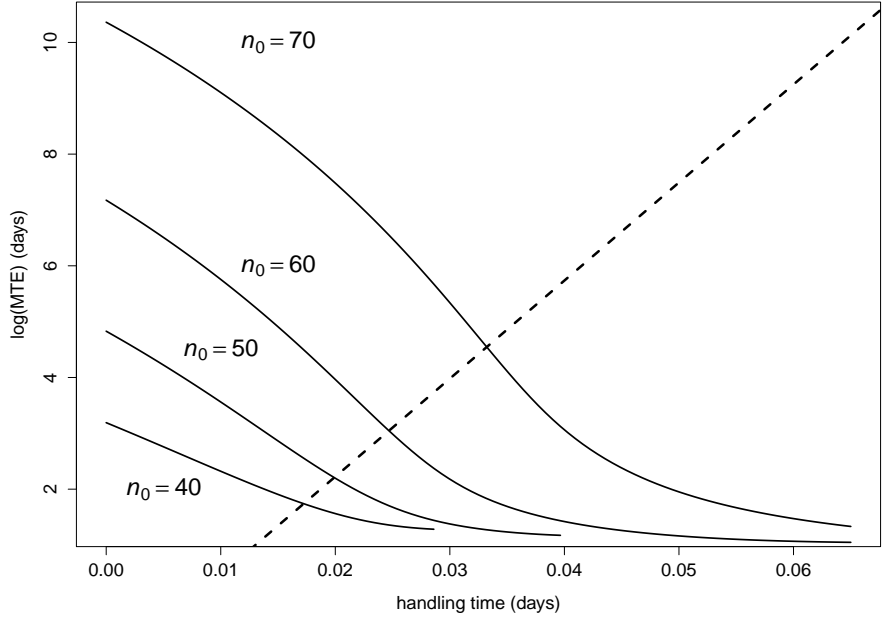}
  \end{center}
\caption{Logarithm of the mean time to extinction (MTE) (when the quasistationary distribution is set as initial condition of equation \eqref{me}) as a function of handling time for different values of the stable fixed point $n_0$, for a type II functional response. The dashed line shows the values of handling time at which there is a change in the stability of the extinction state (left is unstable, right is stable). The curves are drawn for all values of handling time which keep the stable equilibrium density $n_0$ fixed and the corresponding attack rate \eqref{ah} finite i.e., for $h < min(h_1,h_0)$. The values of $h_1$ and $h_0$ are obtained from expressions \eqref{hh} and \eqref{h0}. We used expression \eqref{mte} to obtain the MTE. The growth parameters and the number of predators are as specified in the legend of figure \ref{fig1}.}\label{fig3}
\end{figure*}

\begin{figure*}
  \begin{center}
    \includegraphics[width=0.65\textwidth]{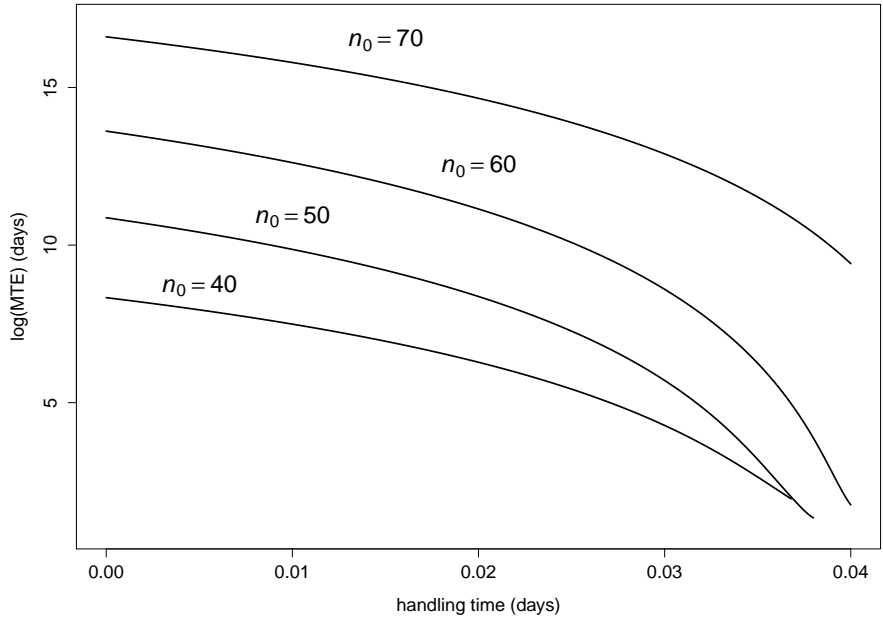}
  \end{center}
\caption{Logarithm of the mean time to extinction (MTE) as a function of handling time for different values of the stable fixed point $n_0$, for type III functional response ($q=1$). We used expression \eqref{mte} to obtain the MTE. The growth parameters and the number of predators are as specified in the legend of figure \ref{fig1}.}\label{fig4}
\end{figure*}

Figures \ref{fig3} and \ref{fig4} show the direct effect of the foraging parameters on the MTE when the equilibrium density is kept fixed. In that case we see that, when the extinction state is unstable, the MTE decreases exponentially with handling time (figure \ref{fig3}), or, in the case of type III functional response, more than exponentially (figure \ref{fig4}). When the extinction state is stable, the MTE decreases less than exponentially (figure \ref{fig3}). We fitted exponential curves to the MTE for four different values of $n_0$, with type II functional response, before the transcritical bifurcation i.e. when the extinction state is unstable. We observe an increase in both the slope and the intercept with increasing carrying capacity\footnote{The values of the slope are fitted using a least square method and are, for different values of the fixed point: $n_0 = 40$, slope -85.8; $n_0 = 50$, slope -133.7; $n_0 = 60$, slope -168.3; $n_0 = 70$ slope -173.}.

The MTE of the model without predation is extremely large ($10^{25} days$), meaning that extinction will never occur in a biological time when there are no predators. We observe a decrease of 20 orders of magnitude in the MTE caused by predation interaction. This decrease is partly as a result of a decrease in the equilibrium population size alone, making extinction due to demographic stochasticity more likely. However demographic stochasticity is important also when the equilibrium density is kept fixed. For example, for a realistic range of foraging parameters, for a type II functional response, at a given fixed point (e.g. $n_0 = 50$) the MTE varies by about 5 orders of magnitude with handling time. The magnitude of this variation increases up to 10 orders of magnitude for increasing values of the stable fixed point (figure \ref{fig3}).

\section{Discussion}

We have shown how different choices of the foraging parameters vary the mean time to extinction by up to 10 orders of magnitude when equilibrium population is kept constant. The effect of the foraging parameters can be intuitively explained by looking at the strength of population regulation at the equilibrium density i.e., the slope of the functional response at the fixed point (Figure \ref{fig1}). Different handling times and attack rates that keep the stable equilibrium fixed result in decreasing maximum saturation rate (increasing handling time) and increasing half saturation density (increasing attack rate). This regulation doesn't alter the actual carrying capacity of the prey population, but can change drastically the demographic effects of predation.
These demographic effects are independent of environmental stochasticity and become relevant for extinction risk only for small populations sizes \citep{Lande2004}. Therefore when the predation interaction is present it is not possible to reliably estimate extinction risk without having a measure of the foraging parameters.

There is a wide literature describing experimental measures of foraging parameters (attack rate, handling time and scaling exponent) in laboratory communities. These studies include predator prey interactions among terrestrial and aquatic organisms such as protists \citep{Hammill2010} and arthropods \citep{Spitze1985, Smout2010}. However similar measurements of the nature of predator prey interactions are absent in most of the studies related to the extinction risk of individual species. The previous focus on individual species characteristics is presumably due to lack of multispecies time series data \citep{Sabo2008}.

We have shown with a simple model how trophic interaction can be relevant in assessing extinction risk of a target species. Our result could be investigated analytically using refined approximation techniques \citep{Ovaskainen2010}. In \citep{Nasell2001} and \citep{Assaf2010} and  an approximate expression of the quasistationary distribution and the mean time to extinction is derived for the stochastic logistic model and the SIS model of epidemics, both slight simplification of our model. The bistability emerging for high values of handling time, for type III functional response, could also be investigated using approximation techniques.

Another natural route for further investigations would be to constrain our analysis of parameter space to combinations of foraging parameters that occur in reality. Introducing allometric relationships between foraging parameters and relating them to the growth parameters would be one way to constrain such analysis. We have shown this dependence as a function of the stable equilibrium density in equation \eqref{ah}. This relation can be generalized using allometric scaling relations between attack rate and handling time \citep{Brose2006}. Such an allometric scaling would require a more general formulation of the model, including predator biomass and prey biomass as other parameters.

This work has application both on studies on extinction risk and on studies on foraging theory and we aim to extend our analysis to multispecies communities in future studies. Most of the existing theoretical studies about complex communities do not incorporate the effects of demographic stochasticity and use deterministic measures of persistence to assess the extinction risk \citep{Hofbauer2008, Dunne2009, Sahasrabudhe2011}. Our findings fit into recent application of stochastic methods to population biology \citep{McKane2004, Ovaskainen2010, Black2012} and can be used to increase the predictive understanding of extinction processes.

For further information and discussion of this paper see \url{http://purl.org/net/extinction}.

\begin{appendices}

\section{Qualitative Behavior} \label{App1}
 
We derive an adimensional formulation of \eqref{det}. We scale the number of individuals with the actual carrying capacity $m = n/k$, and the characteristic time with the intrinsic birth rate $\tau=\lambda t$. With this substitution we have:
\begin{equation}
\frac{dm(\tau)}{d\tau}=m(I-m)-\frac{m^{q+1}}{a+bm^{q+1}}. \label{det3}
\end{equation}

Extinction is always a fixed point of the system. The other fixed points are where the per capita growth term equals the per capita predation rate and are given by the intercepts between the per capita growth term $(I-n)$ and the per capita functional response $[m^q/(a+bm^{q+1})]$ (see figure \ref{fig1}).
In mathematical terms, putting $dm/d\tau = 0$ into equation \ref{det3}, the fixed points of the system are the extinction state $n=0$ and the solutions of the algebraic equation
\begin{equation}
bm^{q+2}-bIm^{q+1}+m^q+am-Ia = 0. \label{eql}
\end{equation}
Where the adimensional parameters have following meanings:
\begin{itemize}
\item $I = 1 - 1 / R_0$ where $R_0 = \lambda/\mu$ is called basic reproductive ratio
\item $a$ is the ratio between intrinsic birth rate and per capita attack rate at the maximum population density of the prey species
\begin{equation}
a = a(q) = \frac{\lambda}{\alpha k^qp}. 
\end{equation}
Note that $a$ is the only adimensional parameter which depends on the scaling exponent $q$. This parameter can be called \emph{inverse adimensional attack rate}.
\item  The last parameter is the one which takes into account handling time $h$
\begin{equation}
b=\lambda h\frac{k}{p}.
\end{equation}
This can be seen as the ratio between handling time and average generation time scaled with the ratio between maximum prey density $k$ and the (fixed) number of predators $p$. This parameter can be called \emph{scaled adimensional handling time}.
\end{itemize}

From the right lower panel of figure \ref{fig1} we see that equation \eqref{eql} has at most 3 real and positive solutions. Note that there will be positive solutions only when $0<R_0<1$ i.e. when $\lambda < \mu$.

As a special case we can look at the solutions of \eqref{eql} when $q=0$, i.e., for type II functional response. Then equation \eqref{eql} simplies to
\begin{equation}
bm^2 + (a-bI)m + (1-aI) = 0. \label{eql2}
\end{equation}
This equation has real solutions when $(a+bI)^2 > 4b$. If this is true then the two real solutions of \eqref{eql2} will be
\begin{equation}
m_{1,2} = \left(\frac{(bI-a) \pm \sqrt{(a+bI)^2-4b}}{2(1-aI)}\right). \label{fix}
\end{equation}
Now we can look at the stability of the fixed points of equation \eqref{det3}, performing the classical stability analysis. We compute the Jacobian of equation \eqref{det3}
\begin{equation}
J(m) = \frac{d}{dm}\left(\frac{dm}{dt}\right)= I - 2m - \frac{(q+1)am^q}{(a+bm^{q+1})^2}, \label{jacob}
\end{equation}
and putting $m=0$ into \eqref{jacob} we have 
\begin{itemize}
\item if $q=0$ then $J(0) = r - \alpha p$ where $r=\lambda - \mu$ so
\begin{enumerate}
\item if $r<\alpha p$ the extinction state is stable
\item if $r>\alpha p$ the extinction state is unstable
\end{enumerate}
\item if $q > 0$ the extinction state is always unstable
\end{itemize}
The stability of the extinction state gives raise to the the two extinction scenarios described in the Methods section.

\section{Mean time to Extinction} \label{App2}

In this appendix we present the mathematical and numerical tools needed to obtain the quasi-stationary distribution and the mean time to extinction of a general birth and death process when ultimate extinction is certain. We define the process as the time evolution of a random variable $\{X(t),t\geq 0\}$ in a finite space state $\{0,1,\cdots ,k\}$ where the origin is an absorbing barrier \citep{Nisbet1982}.
We recall the master equation \eqref{me} and write it in a more compact way using a vectorial notation:
\begin{equation}
\frac{d{\bf p}(t)}{dt}={\bf p}(t){\bf A}, \label{me2}
\end{equation}
where ${\bf p}(t)=(p(0,t),p(1,t),\cdots,p(k,t))$ is the row vector containing the state probabilities and the matrix ${\bf A}$ contains the transition rates as follows:
\begin{equation}
{\bf A} = \left(\begin{array}{ccccc}-G(0)&B(0)&0&\cdots&0\\\ D(1)&-G(1)&B(1)&\cdots&0\\0&D(2)&-G(2)&\cdots&0\\\cdots&\cdots&\cdots&\cdots&\cdots\\0&0&0&\cdots&-G(k)\end{array} \right),
\end{equation}
with $G(n) = B(n)+D(n)$. {\bf A} is a tridiagonal matrix with all row sums equals to $0$. Note also that, using the rates \eqref{BD} the first row is a row of zeros.
 The solution of the master equation \eqref{me} will give the probability of having $n$ individuals at time $t$; in other words $p(n,t)=P\{X(t)=n\}$. Once the birth and death process is defined we derive two auxiliary processes\citep{Nasell2001} $\{X^{(0)}(t)\}$ and $\{X^{(1)}(t)\}$ associated to $\{X(t)\}$, both with reduced state space $\{1,2,\cdots,k\}$:
\begin{itemize}
\item The first process $\{X^{(0)}(t)\}$ is equal to $\{X(t)\}$ but with reduced state space. So the rates of transition are not changed except $D(1)$:
\begin{equation}
\begin{array}{ccc}
&B^{(0)}(n)=B(n), \\
&D^{(0)}(n)=D(n),\\
&D^{(0)}(1)=0.
\end{array}
\end{equation}
But the assumption on $D^{(0)}(1)$ ensures that there will never be extinction, in other words that there is no absorbing state.
\item The second process $\{X^{(1)}(t)\}$ is defined, in the reduced state space, with the following transition rates:
\begin{equation}
\begin{array}{cc}
&B^{(1)}(n)=B(n), \\
&D^{(1)}(n)=D(n-1), \\
&D^{(1)}(0) = 0,
\end{array}
\end{equation}
and corresponds to a population in which there is always at least one ``immortal" individual.
\end{itemize}
It is important to underline that the two auxiliary processes we have defined have no ecological meaning. $\{X^{(0)}(t)\}$ and $\{X^{(1)}(t)\}$ will be used only as mathematical tools in order to obtain information on the time to extinction of the birth-death process \eqref{bd}.

The stationary distributions of the two auxiliary processes are:
\begin{equation}
p^{(0)}(n)=\frac{T(n)}{\sum_{n=1}^k T(n)},      p^{(1)}(n)=\frac{R(n)}{\sum_{n=1}^k R(n)},
\end{equation}
where $T(n)$ and $R(n)$ are defined as follows:
\begin{equation}
\begin{split}
T(n)&=\frac{B(1)B(2)\cdots B(n-1)}{D(2)D(3)\cdots D(n)},\\
 R(n)&=\frac{B(1)B(2)\cdots B(n-1)}{D(1)D(2)\cdots D(n-1)}.
\end{split}
\end{equation}
Note that $T(n)=R(n)\frac{D(1)}{D(n)}$.
 
Now we can partition the state space of the original process into two subsets, $\{0\}$ and $Q=\{1,2,\cdots,k\}$. $Q$ is the set o transients for $\{X(t)\}$ and the state space for the two auxiliary processes while $\{0\}$ is the absorbing state for $\{X(t)\}$. Correspondingly we can partition the state vector ${\bf p}(t)$ and the transition matrix ${\bf A}$, and obtain from equation \eqref{me2}
\begin{equation}
\left[\frac{dp(0,t)}{dt};\frac{d{\bf p}_Q(t)}{dt}\right]=[p(0,t);{\bf p}_Q(t)]\left(\begin{array}{cc}0&{\bf 0}\\{\bf a}&{\bf A_Q}\end{array}\right).
\end{equation}
Here ${\bf p}_Q(t)$ is the vector of probabilities in the transient states and ${\bf a}=(D(1),0,\cdots,0)$. With this separation we can split the master equation \eqref{me2} into:
\begin{equation}
\begin{array}{cc}
&dp(0,t)/dt={\bf p}_Q(t){\bf a}=D(1)p(1,t),\\\\
&d{\bf p}_Q(t)/dt={\bf p}_Q(t){\bf A_Q}.
\end{array}\label{split}
\end{equation}
Before absorption the process takes values in the set of the transients.

As in equation \eqref{pc} we define the conditional probability $p_c(n,t)=P\{X(t)=n|X(t)>0\}$ of having $n$ individuals at time $t$ knowing absorption has not occurred yet. And, using equations \eqref{split}, we can express it in vectorial form:
\begin{equation}
{\bf p}_c(t)=\frac{{\bf p}(t)}{1-p(0,t)}=\frac{{\bf p}_Q(t)}{1-p(0,t)}. \label{cond}
\end{equation}
Differentiating equation \eqref{cond} and using the master equation \eqref{me} and equations \eqref{split} we obtain an equation for ${\bf p}_c(t)$:
\begin{equation}
\begin{split}
\frac{d{\bf p}_c(t)}{dt}&=\frac{d{\bf p}_Q}{dt}\left(\frac{1}{1-p(0,t)}\right)+\frac{{\bf p}_Q(t)}{(1-p(0,t))^2}\frac{dp(0,t)}{dt}\\
&={\bf p}_c(t){\bf A_Q}+D(1)p_c(1,t){\bf p}_c(t). \label{quasi}
\end{split}
\end{equation}
Putting to zero the right-hand side of expression \eqref{quasi} we obtain an equation for the quasi-stationary distribution $\bm{\pi}= (\pi(1),\pi(2),\cdots ,\pi(k))$, defined as the distribution of the transient states conditioned on the fact that there has been no extinction yet:
\begin{equation}
{\bm \pi}{\bf A_Q}=-D(1)\pi(1) {\bm \pi}.
\end{equation}
In other words the quasistationary distribution ${\bm \pi}$ is the left eigenvector of ${\bf A}_Q$ with eigenvalue $-D(1)\pi(1)$.

It can be shown \citep{Nasell2001} using a recursive solution of the master equation \eqref{quasi}, that the $\pi(n)$ satisfy the recursive formula:
\begin{equation}
\pi(n)=T(n)\sum_{i=1}^n\frac{(1-\sum_{j=1}^{i-1}\pi(j))}{R(i)}\pi(1). \label{recursive}
\end{equation}
Once $\pi(1)$ is known then $\pi(2), \pi(3), \dots, \pi(n)$ can be determined iteratively. But $\pi(1)$ can only be obtained knowing all the other elements by the relation $\sum_n \pi(n)=1$. For this reason the analytic determination of ${\bm \pi}$ is limited to birth death processes with linear transition rates and this is not our case. However there is an iterative method that can be used to derive numerical approximations for the quasistationary distribution of our process:

\begin{itemize}
\item Start with an initial guess for $\pi(1)$.
\item Obtain all the $\pi(n)$ using the \eqref{recursive} and compute $S=\sum_n\pi(n)$.
\item Start the iteration again with $\pi^{I}(1)=\pi(1)/S$ and obtain all the $\pi^{I}(n)$.
\item Repeat the process until $\Vert \pi^{K+1}(n)-\pi^K(n)\Vert<\delta$. The value $\delta$ gives the precision of the algorithm.
\end{itemize}

The time to extinction\footnote{Be careful this $\tau$ is different from the adimensional time used in appendix A} $\tau$ is a random variable that depends on the initial distribution of the process \citep{Nasell2001}. We call $\tau_Q$ the time to extinction of the birth death process when the quasi stationary distribution ${\bm \pi}$ is set as initial distribution. And we call $\tau_n$ the time to extinction when the initial condition is $X(0)=n$ i.e., when $p(n,0)=1$.

If absorption has occurred at time $t$ then the events $\{\tau < t\}$ and $\{X(t)=0\}$ are identical:
\begin{equation}
P\{\tau < t\}=P\{X(t)=0\}=p(0,t).
\end{equation}
Once the quasistationary distribution is known then the MTE\footnote{note that with our notation $MTE = E(\tau_Q)$.} is given by expression \eqref{mte}.

The explicit expression of time to extinction with an arbitrary initial condition is more difficult to obtain. It is a standard result for birth death processes theory \citep{Nisbet1982} that at least the expectation can be determined explicitly when $X(0)=n$:
\begin{equation}
E(\tau_n)=\frac{1}{D(1)}\sum_{i=1}^n\frac{1}{R(i)}\sum_{j=i}^k T(j). \label{etn}
\end{equation}
Note that putting $n=1$ in the above formula the expected time to extinction from the state 1 can be written as follows:
\begin{equation}
E(\tau_1)=\frac{1}{D(1)}\sum_{j=1}^k T(j)=\frac{1}{D(1) p^{(0)}(1)}.
\end{equation}
The expected time to extinction from a state $n$ can therefore be written in an alternative form in function of the stationary distribution of the first auxiliary process:
\begin{equation}
E(\tau_n)=E(\tau_1)\sum_{i=1}^n\frac{1}{R(i)}\sum_{j=i}^k p(j)^{(0)}.
\end{equation}
Then the expected time to extinction for an arbitrary initial distribution $\{p(n,0)\}$ can be derived from \eqref{etn}:
\begin{equation}
E(\tau)=\frac{1}{D(1)}\sum_{j=1}^k T(j)\sum_{i=1}^j\frac{1}{R(i)}\sum_{n=i}^k p(n,0),
\end{equation}
with the assumption that the initial distribution is supported on the set of the transient states.

\end{appendices}

\section*{Acknowledgments}

GMP would like to thank Antonio Scala and Rich Williams for helpful discussions. The project is funded by Microsoft Research Limited and by the University of Zurich.

\bibliographystyle{model2-names}

\end{document}